\documentstyle[twocolumn]{article}
\begin{document}

\noindent
{\large {\bf A Model of Magnetic Monopoles}}

\vspace{0.5 cm}

\noindent
Rainer W. K\"uhne \\
{\it Fachbereich Physik, Universit\"at Wuppertal, 42097 Wuppertal, Germany \\
e-mail: kuehne@theorie.physik.uni-wuppertal.de}

\vspace{0.5 cm}

\noindent
The possibility of the existence of magnetic charges is one of the 
greatest unsolved issues of the physics of this century. 
The concept of magnetic monopoles has at least two attractive features: 
(i) Electric and magnetic fields can be described equivalently. 
(ii) In contrast to quantum electrodynamics models of monopoles are able 
to explain the quantization of electric charge. 
We suggest a quantum field theoretical model of the electromagnetic 
interaction that describes electricity and magnetism as equivalent as 
possible. This model requires the cross-section of Salam's ``magnetic 
photon'' to depend on the absolute motion of the electric charge with 
which it interacts. We suggest a tabletop experiment to verify this 
magnetic photon. Its discovery by the predicted effect would have 
far-reaching consequences: (i) Evidence for a 
new gauge boson and a new kind of radiation which may find applications 
in medicine. (ii) Evidence for symmetrized Maxwell equations. (iii) 
Evidence for an absolute rest frame that gives rise to local physical 
effects and violation of Einstein's relativity principle.

\vspace{0.5 cm}


\vspace{0.5 cm}

\noindent
The quantization of electric charge 
is well-known since the discovery of the proton in 1919. This remarkable 
observation remained unexplained.

Further quantized charges have been established. The group $SU(2)_{w}$ of 
the weak interaction explains the quantization of isospin, and the group 
$SU(3)_{c}$ of the strong interaction explains the quantization of colour 
charge.

For this reason we propose the ``analogy postulate'': 
``The quantization of electric charge results from the underlying group 
structure of the electromagnetic interaction.'' 
Hence, we will require neither quantum gravity (electric charge as a 
topological quantum number), nor spontaneous symmetry breaking (monopoles 
of soliton type$^{1,2}$), nor unification with other forces (charge 
quantization resulting from the group structure underlying grand unified 
theories).

The electromagnetic angular momentum generated by the Lorentz force in a 
system consisting of a magnetic monopole and an electric charge is 
independent of their separation$^{3}$. Angular momentum is quantized in 
units of $\hbar /2$, where $\hbar =h/2\pi$ denotes Planck's constant. 
This condition can be satisfied only if both electric 
and magnetic charge are quantized$^{4}$. This is the famous Dirac 
quantization condition $eg=h$, where $e$ and $g$ denote unit 
electric and unit magnetic charge.

Magnetic monopoles were discussed long before this finding. 
The motivation was to describe electric and magnetic fields equivalently 
by symmetrized Maxwell equations. We will elevate this to the 
``symmetry postulate'': ``The fundamental equations of the electromagnetic 
interaction describe electric and magnetic charges, electric and magnetic 
field strengths, and electric and magnetic potentials equivalently.''

Dirac$^{4}$ was the first to write down these symmetrized Maxwell equations. 
Let 
$J^{\mu}=(P, {\bf J})$ denote the electric four-current and 
$j^{\mu}=(\rho , {\bf j})$ the magnetic four-current. Furthermore, we require  
the well-known four-potential $A^{\mu}=(\Phi , {\bf A})$ and a new 
four-potential 
$a^{\mu}=(\varphi , {\bf a})$. Expressed in three-vectors Dirac's symmetrized 
Maxwell equations read,
\begin{eqnarray}
\nabla\cdot {\bf E} & = & P \\
\nabla\cdot {\bf B} & = & \rho \\
\nabla\times {\bf E} & = & - {\bf j} - \partial_{t} {\bf B} \\
\nabla\times {\bf B} & = & + {\bf J} + \partial_{t} {\bf E}
\end{eqnarray}
and the relations between field strengths and potentials are
\begin{eqnarray}
{\bf E} & = & - \nabla\Phi - \partial_{t} {\bf A} -\nabla\times {\bf a} \\
{\bf B} & = & - \nabla\varphi - \partial_{t} {\bf a} +\nabla\times {\bf A}.
\end{eqnarray}

The second four-potential is required not only by the symmetry postulate, 
but also by the proven impossibility to construct a manifestly covariant 
one-potential model 
of quantum electromagnetodynamics$^{5-8}$.

Although only one of the suggested two-potential models$^{4, 9-15}$ 
explicitely states the possibility of the existence of a ``magnetic photon'' 
(ref. 15), the other two-potential models were eventually considered as 
two-photon models$^{16}$.

Any viable two-photon concept of magnetic monopoles has to satisfy the 
following conditions.

(i) In the absence of both magnetic charges and the magnetic photon field, 
the model has to regain the $U(1)$ gauge symmetry of quantum 
electrodynamics.

(ii) In the absence of both electric 
charges and the photon field, the symmetry postulate requires the model to 
yield the $U'(1)$ gauge symmetry of ``quantum magnetodynamics''.

(iii) The gauge group has to be Abelian, because the photon carries neither 
electric nor magnetic charge. Because of the symmetry postulate also the 
magnetic photon has to be neutral.

(iv) The gauge group may not be simple, because quantum 
electromagnetodynamics includes the two coupling constants 
$\alpha_{E}=e^{2}/4\pi\simeq 1/137$ and $\alpha_{M}=g^{2}/4\pi =
1/4\alpha_{E}\simeq 34$.

The only gauge group that satisfies these four conditions is the group 
$U(1)\times U'(1)$.

A two-photon model has already been suggested by Salam$^{15}$. 
According to his model the photon couples via vector coupling with leptons 
and hadrons, but not with monopoles. The magnetic photon couples via 
vector coupling with monopoles and via tensor coupling with hadrons, but not 
with leptons.

This model came under severe criticism. Although positron and proton have 
the same electric charge and no magnetic charge, the model can discriminate 
them (i. e. leptons and hadrons). For this reason Salam's model does not 
generate the Lorentz force between electric charge and monopole. As a 
consequence, it does not satisfy the powerful Dirac quantization 
condition. For this reason Salam's model was rejected by Taylor$^{16}$.

The problem raised by Taylor can be overcome by the following argumentation. 
Salam considered the tensor coupling of the hadron-monopole system as 
derivative coupling. This kind of coupling is well-known from meson theory 
where vector mesons are able to interact with baryons via both vector and 
tensor coupling. However, derivative coupling is possible only where the 
particles are composite. Hence, Salam's model includes no 
interaction between lepton and magnetic photon. -- We emphasize the 
correctness of the interpretation of tensor coupling as derivative coupling 
in meson theory.

To generate the Lorentz force between electric and magnetic charges we 
have to introduce a new kind of tensor coupling. This is required also, 
because here we have two kinds of interacting charges (electric and 
magnetic).

The Coulomb force between two (unit) electric charges is 
$e^{2}/4\pi r^{2}$. Because of the symmetry postulate the magnetic force 
between two (unit) magnetic charges is $g^{2}/4\pi r^{2}$. And the 
Lorentz force between (unit) electric and (unit) magnetic charge is 
$egv/4\pi r^{2}$, where $v$ denotes the relative velocity of the two 
charges.

This suggests the introduction of ``velocity coupling'':

(i) The photon couples via vector coupling with electric charges.

(ii) The magnetic photon couples via vector coupling with magnetic charges.

(iii) The photon couples via tensor coupling with magnetic charges. In 
contrast to meson theory, however, the $u^{\mu}$ of tensor coupling, 
$\sigma^{\mu\nu}u_{\nu}$, has to be interpreted as a four-velocity 
(``velocity coupling'').

(iv) The magnetic photon couples via tensor coupling (interpreted as 
velocity coupling instead of derivative coupling) with electric charges.

In the case of the interacting 
monopole-electric charge system the exchanged boson (either photon or 
magnetic photon) is virtual and the four-velocity of velocity coupling 
is the relative four-velocity between the charges.

Charged quanta are required to emit and absorb the same bosons as real 
(on-mass-shell) particles as those virtual (off-mass-shell) bosons via 
whom they interact with other charged quanta. This is because the Feynman 
rules are symmetric with respect to virtual and real particles.

In the case of emission and absorption reactions of real bosons, $u^{\mu}$ 
cannot be interpreted as a relative four velocity between charged 
quanta in the initial state, as there is only one charged quantum present. 
As a consequence, 
$u^{\mu}$ has to be interpreted as the absolute four-velocity of the initial 
charged quantum.

In contrast to general belief an absolute rest frame is not forbidden. 
Instead, a number of reasons support its existence.

(i) General relativity gives rise to an expanding universe and therefore to a 
finite-sized light zone. The center-of-mass frame of this Hubble sphere can be 
regarded as a preferred frame.

(ii) According to Bondi and Gold$^{17}$ a preferred motion is given at each 
point of space by cosmological 
observations, namely the redshift-distance relation generated by the 
superposition of the Hubble and the Doppler effect which is isotropic only 
for a unique rest frame. 

(iii) The dipole anisotropy of the cosmic 
microwave background radiation due to the Doppler effect by the Earth's 
motion was predicted in accordance with Lorentz invariance$^{18, 19}$. 
This ``aether drift'' of the Earth was later discovered$^{20}$ and 
measured to be $v_{earth}=370$ km/s ($\pm 30$ km/s, because of the Earth's 
motion around the Sun).

Now we have the tools to construct the Lagrangian for a spin 1/2 fermion field 
$\Psi $ of rest mass $m_{0}$, electric charge 
$Q$, and magnetic charge $q$ within an electromagnetic field. 
By using the tensors
\begin{eqnarray}
F^{\mu\nu} & \equiv & \partial^{\mu}A^{\nu}- \partial^{\nu}A^{\mu} \\
f^{\mu\nu} & \equiv & \partial^{\mu}a^{\nu}- \partial^{\nu}a^{\mu}
\end{eqnarray}
the Lagrangian of the Dirac fermion within the electromagnetic field reads, 
\begin{eqnarray}
{\cal L} & = & - \frac{1}{4}F_{\mu\nu}F^{\mu\nu}
               - \frac{1}{4}f_{\mu\nu}f^{\mu\nu} \nonumber \\
 & &  + \bar\Psi i\gamma^{\mu}\partial_{\mu}\Psi - m_{0}\bar\Psi \Psi 
 \nonumber \\
 & &  -Q\bar\Psi \gamma^{\mu}\Psi A_{\mu} - q\bar\Psi \gamma^{\mu}\Psi a_{\mu}
 \nonumber \\ 
 & &  +Q\bar\Psi \gamma^{5}\sigma^{\mu\nu}u_{\nu}\Psi a_{\mu} \nonumber \\
 & &  +q\bar\Psi \gamma^{5}\sigma^{\mu\nu}u_{\nu}\Psi A_{\mu}.
\end{eqnarray}
By using the Euler-Lagrange equations we obtain the Dirac equation
\begin{eqnarray}
(i\gamma^{\mu}\partial_{\mu}-m_{0})\Psi & = & (Q\gamma^{\mu}A_{\mu}
+q\gamma^{\mu}a_{\mu} \nonumber \\
 & & -Q\gamma^{5}\sigma^{\mu\nu}u_{\nu}a_{\mu} \nonumber \\
 & & -q\gamma^{5}\sigma^{\mu\nu}u_{\nu}A_{\mu})\Psi .
\end{eqnarray}
By introducing the four-currents
\begin{eqnarray}
J^{\mu} & = & Q\bar\Psi \gamma^{\mu}\Psi -q\bar\Psi \gamma^{5}\sigma^{\mu\nu}
u_{\nu}\Psi \\
j^{\mu} & = & q\bar\Psi \gamma^{\mu}\Psi -Q\bar\Psi \gamma^{5}\sigma^{\mu\nu}
u_{\nu}\Psi
\end{eqnarray}
the Euler-Lagrange equations yield the two Maxwell equations
\begin{eqnarray}
J^{\mu} & = & \partial_{\nu}F^{\nu\mu} = \partial^{2}A^{\mu} 
- \partial^{\mu}\partial^{\nu}A_{\nu} \\
j^{\mu} & = & \partial_{\nu}f^{\nu\mu} = \partial^{2}a^{\mu} 
- \partial^{\mu}\partial^{\nu}a_{\nu}.
\end{eqnarray}
Evidently, the two Maxwell equations are invariant under the 
$U(1)\times U'(1)$ gauge transformations
\begin{eqnarray}
A^{\mu} & \rightarrow & A^{\mu}-\partial^{\mu}\Lambda \\
a^{\mu} & \rightarrow & a^{\mu}-\partial^{\mu}\lambda .
\end{eqnarray}
Furthermore, the four-currents satisfy the continuity equations
\begin{equation}
0=\partial_{\mu}J^{\mu}= \partial_{\mu}j^{\mu}.
\end{equation}
The electric and magnetic field are related to the tensors above by
\begin{eqnarray}
E^{i} & = & F^{i0}- \frac{1}{2}\varepsilon^{ijk}f_{jk} \\
B^{i} & = & f^{i0}+ \frac{1}{2}\varepsilon^{ijk}F_{jk}.
\end{eqnarray}
Finally, the Lorentz force is
\begin{eqnarray}
K^{\mu} & = & Q(F^{\mu\nu}+ \frac{1}{2}\varepsilon^{\mu\nu\varrho\sigma}
              f_{\varrho\sigma})u_{\nu} \nonumber \\
 & & + q(f^{\mu\nu}+ \frac{1}{2}\varepsilon^{\mu\nu\varrho\sigma}
              F_{\varrho\sigma})u_{\nu},
\end{eqnarray}
where $\varepsilon^{\mu\nu\varrho\sigma}$ denotes the totally 
antisymmetric tensor.

This model does not contain any free parameters. Hence, it allows clear and 
decisive predictions for its verification.

The absolute rest frame predicted above gives rise to 
local physical effects. In a terrestrial 
laboratory the interaction cross section of a real magnetic photon (with 
ordinary matter in the terrestrial rest frame) is predicted to be 
smaller than the one of a real photon of the same energy, the suppression 
factor is $(v_{earth}/c)^{2}\simeq 
1.5\times 10^{-6}$. Hence, each reaction that produces photons 
does also create magnetic 
photons. Furthermore, magnetic photons are $7\times 10^{5}$ times harder 
to create, to shield and to absorb 
than photons of the same energy.

A relatively simple experiment to verify this idea is to illuminate a metal 
foil of the 
diameter $(1\ldots 100)\mu$m  by a laser beam and to place a detector 
(photographic plate, charge coupled device, or photomultiplier tube) 
behind the 
foil to measure the intensity of the penetrating fraction of the beam.

The $U(1)\times U'(1)$ model predicts that in contrast to the photons 
majority of the magnetic photons 
penetrate the foil. This is because the penetration depth of photons 
of the visible light in metals is a 
few nanometers, whereas the one of magnetic photons is predicted to be a 
few millimeters. 
The detected intensity is predicted to be about $10^{-12}$ times the one 
of the 
original laser beam (depending on the efficiency of the detector, usually of 
order one for photons and as a 
consequence of order $10^{-6}$ for magnetic photons).

It will be of great benefit to perform the proposed experiment. 
The discovery of the magnetic photon by the predicted effect would have 
far-reaching consequences:

(i) Evidence for a new vector gauge boson and a new kind of radiation 
(``magnetic photon rays'').

(ii) Indirect evidence for magnetic monopoles and therefore the explanation 
of the quantization of electric charge.

(iii) Evidence that the fundamental equations of the electromagnetic 
interaction are symmetrized Maxwell equations that are invariant 
under $U(1)\times U'(1)$ gauge transformations.

(iv) Evidence that the dipole anisotropies of both the cosmic microwave 
background and the cosmological redshift distribution (by the Hubble effect 
and the Doppler effect)  
define an absolute rest frame and that it gives rise to local physical 
effects. This means a violation of the relativity principle.

Rays of magnetic photons may find industrial 
applications. Their 
penetration depth in matter is predicted to be $7\times 10^{5}$ times the 
one of photon rays of the 
same wavelength. As a consequence, rays of magnetic photons may become an 
appropriate means for 
testing of materials (to detect material errors). For the same reason, 
they may find applications 
in those clinical purposes which are presently the domains of X-ray, 
holograph and ultrasonic diagnostics.

It remains the task for further studies to examine the following issues. 
(i) The mass of the lightest 
magnetic monopole has to be calculated. According to the most simple 
approach it can be estimated to be 
$m_{e}g^{2}/e^{2}\simeq 2$~GeV~, where $m_{e}$ is the electron mass$^{21}$. 
(ii) The coupling 
constant of the monopole-monopole interaction is as large as 
$1/4\alpha_{E}\simeq 34$. It has to be 
examined how the arising difficulties that concern unitarity and 
non-locality can be solved. - A 
similar problem arose in meson theory. The coupling constant of the pion 
is as large as 14.4~. The 
arising problems were solved by the quark substructure of hadrons. 
(iii) Einstein$^{22}$ rejected the ``luminiferous aether'', because 
special relativity does not require an absolute rest frame. It would 
be interesting to learn whether aether is possible in general relativity. 
(iv) The $U(1)\times U'(1)$ model has to be embedded in a Grand Unified 
theory. One may imagine the underlying group to be 
$SU(5)\times SU'(5)$, where the primed group describes the hypothetical 
magnetic photon, (chromo-)magnetic 
gluons and (iso-)magnetic W, Z, X, and Y bosons.

\vspace{0.5 cm}

\noindent
1. G. t'Hooft, {\it Nucl. Phys. B} {\bf 79}, 276 (1974). \\
2. A. M. Polyakov, {\it Sov. Phys. JETP Lett.} {\bf 20}, 194 (1974). \\
3. J. J. Thomson, {\it Elements of the Mathematical Theory of Electricity 
   and Magnetism} (Cambridge Univ. Press, 1904). \\
4. P. A. M. Dirac, {\it Proc. Roy. Soc. A} {\bf 133}, 60 (1931). \\
5. D. Zwanziger, {\it Phys. Rev. B} {\bf 137}, 647 (1965). \\
6. S. Weinberg, {\it Phys. Rev. B} {\bf 138}, 988 (1965). \\
7. C. R. Hagen, {\it Phys. Rev. B} {\bf 140}, 804 (1965). \\
8. A. S. Goldhaber, {\it Phys. Rev. B} {\bf 140}, 1407 (1965). \\
9. N. Cabibbo and E. Ferrari, {\it Nuovo Cim.} {\bf 23}, 1147 (1962). \\
10. J. S. Schwinger, {\it Phys. Rev.} {\bf 144}, 1087 (1966). \\
11. J. S. Schwinger, {\it Phys. Rev.} {\bf 151}, 1048 (1966). \\
12. J. S. Schwinger, {\it Phys. Rev.} {\bf 151}, 1055 (1966). \\
13. A. Rabl, {\it Phys. Rev.} {\bf 179}, 1363 (1969). \\
14. D. Zwanziger, {\it Phys. Rev. D} {\bf 3}, 880 (1971). \\
15. A. Salam, {\it Phys. Lett.} {\bf 22}, 683 (1966). \\
16. J. G. Taylor, {\it Phys. Rev. Lett.} {\bf 18}, 713 (1967). \\
17. H. Bondi and T. Gold, {\it Nature} {\bf 169}, 146 (1952). \\
18. R. N. Bracewell and E. K. Conklin, {\it Nature} {\bf 219}, 1343 (1968). \\
19. P. J. E. Peebles and D. T. Wilkinson, {\it Phys. Rev.} 
    {\bf 174}, 2168 (1968). \\
20. G. F. Smoot, M. V. Gorenstein, and R. A. Muller,
    {\it Phys. Rev. Lett.} {\bf 39}, 898 (1977). \\
21. W. C. Carithers, R. Stefanski, and R. K. Adair, 
    {\it Phys. Rev.} {\bf 149}, 1070 (1966). \\
22. A. Einstein, {\it Ann. Phys. (Leipzig)} {\bf 17}, 891 (1905).

\end{document}